\documentclass[12pt,letterpaper,oneside]{article}
\usepackage[margin=1in]{geometry}
\usepackage{authblk}
\usepackage{natbib}
\usepackage{amsmath, amssymb, hyperref, ulem}
\usepackage{graphicx,subfigure}

\title{Accelerating Time-Reversal Imaging with Neural Operators for Real-time Earthquake Locations}

\author[1]{Hongyu Sun \thanks{Corresponding author: hongyu-sun@outlook.com}}
\author[1]{Yan Yang}
\author[2]{Kamyar Azizzadenesheli}
\author[1]{Robert W. Clayton}
\author[1]{Zachary E. Ross}

\affil[1]{Seismological Laboratory, California Institute of Technology, 1200 E. California Blvd., Pasadena, CA 91125}
\affil[2]{Nvidia Corporation, 2788 San Tomas Expressway,
Santa Clara, CA 95051}

\date{October 2022}

\begin{document}

\maketitle

\begin{abstract}
Earthquake hypocenters form the basis for a wide array of seismological analyses. Pick-based earthquake location workflows rely on the accuracy of phase pickers and may be biased when dealing with complex earthquake sequences in heterogeneous media. Time-reversal imaging of passive seismic sources with the cross-correlation imaging condition has potential for earthquake location with high accuracy and high resolution, but carries a large computational cost. Here we present an alternative deep-learning approach for earthquake location by combining the benefits of neural operators for wave propagation and time reversal imaging with multi-station waveform recordings. A U-shaped neural operator is trained to propagate seismic waves with various source time functions and thus can predict a backpropagated wavefield for each station in negligible time. These wavefields can either be stacked or correlated to locate earthquakes from the resulting source images. Compared with other waveform-based deep-learning location methods, time reversal imaging accounts for physical laws of wave propagation and is expected to achieve accurate earthquake location. We demonstrate the method with the 2D acoustic wave equation on both synthetic and field data. The results show that our method can efficiently obtain high resolution and high accuracy correlation-based time reversal imaging of earthquake sources. Moreover, our approach is adaptable to the number and geometry of seismic stations, which opens new strategies for real-time earthquake location and monitoring with dense seismic networks.

\end{abstract}
\textbf{keywords}: Time reversal imaging, neural operators, earthquake location, wave propagation, correlation

\section{Introduction}

Source locations are important for earthquake monitoring and early warning \citep{zhang2021real}, understanding faulting properties and initiation of earthquake sequences \citep{ross20203d}, and hazard assessment of induced seismicity during industrial injection \citep{lin2020source}. Routine earthquake location workflows usually include phase detection, picking, association, and location. With such a sequentially-staged workflow, the resulting source locations heavily depend on the initial picking results. However, picking is usually done on seismograms from each station separately and may have difficulty when the first arrivals overlap with the coda of a larger event. Furthermore, only P and S arrivals are used in the estimation of earthquake locations, and converted waves that are common in complex media do not contribute to the workflow. Instead, phase pickers may erroneously consider converted waves as first arrivals and introduce errors to the location.

With the availability of dense seismic networks and distributed acoustic sensing, we can directly locate earthquake sources from full waveforms with time reversal imaging and benefit from the coherency of phases between stations. With full waveforms of seismograms instead of picked arrivals, time-reversal imaging has shown its power for revealing earthquake locations and source mechanisms at local \citep{zhu2019hybrid}, regional \citep{mcmechan1985imaging,larmat2008time}, and global \citep{larmat2006time} scales. By backpropagating time-reversed seismograms at receiver locations, seismic-wave energies will refocus at the origins of seismic events, provided with reasonably accurate velocity models and assumptions of non dissipative media \citep{mcmechan1982determination, gajewski2005reverse, artman2010source}. We can distinguish locations and estimate their origin time by detecting the maximum intensity or reasonable focusing in the resulting source images.

The demanding computational cost and sometimes low spatial imaging resolution hinder the universal application of time reversal imaging from earthquake location. Conventional time reversal imaging methods simultaneously backpropagate entire seismograms and thus result in source images by implicitly stacking wavefields at all stations. However, such resulting source imaging generally suffers from low imaging resolution, which makes it challenging to search for source locations from records with low signal-to-noise ratio (SNR). Most recently, \cite{sun2015investigating,nakata2016reverse} propose a cross-correlation imaging method to enhance the spatial resolution of time-reversal source imaging by individually extrapolating wavefields at each station and then cross-correlating these wavefields. However, the computational cost increased by this method is proportional to the number of stations, which imposes challenges for real-time earthquake location with a dense seismic network. To reduce the computational cost of wavefield extrapolation, researchers have grouped several receivers and performed backward wave propagation for each group before cross-correlation \citep{zhu2019hybrid,lin2020source,wu2022crosscorrelation}; however, this comes at a cost of reduced image resolution, and the choice of grouping strategy only depends on experience and is not straightforward \citep{bai2022receiver}. In addition, \cite{baker2005real} directly image earthquake source locations with Kirchhoff reconstruction of ground motions. \cite{li2020efficient} approximate wave-equation solutions with simplified Gaussian beam for efficient source location with time reversal methods. 

Deep learning has become state of the art for most earthquake monitoring tasks, leading to advances in our knowledge about the Earth \citep{ross20203d,li2021basal,yang2022toward}. On one hand, inspired by traditional pick-based earthquake location workflows, deep learning has been used in each step of the sequential workflow \citep{zhang2022loc}, including earthquake detection \citep{ross2018p}, phase picking \citep{ross2018generalized,zhu2019phasenet}, phase association \citep{ross2019phaselink,zhu2022earthquake}, and location \citep{smith2022hyposvi}. On the other hand, various deep learning methods are developed to infer earthquake locations directly from continuous waveforms \citep{van2020automated,zhang2021real,munchmeyer2021earthquake,shen2021array}. By introducing time reversal imaging to deep learning, we propose to combine physical laws of wave propagation with deep neural operators and to determine source locations directly from waveforms recorded at all stations in a seismic network with relatively high accuracy.

In this work, we propose to achieve time-reversal imaging with neural operators for real-time earthquake location. Neural operators \citep{kovachki2021neural} are a generalization of neural networks to learn operators that map between infinite dimensional function spaces. It has been used to solve various partial differential equations (PDEs) with state-of-the-art efficiency \citep{li2020fourier}. In seismology, \citet{yang2021seismic, yang2022accelerated} used Fourier neural operators \citep{li2020fourier} to solve the acoustic and elastic wave equations in 2D and demonstrate their ability to accelerate forward modeling for full-waveform inversion. Here, we train a U-shaped neural operator \citep{rahman2022u} with random source time functions. This allows for the neural operator to learn backward wave propagation and predict time reversed wavefields with negligible compute time. The predicted wavefields can be either stacked or correlated; they result in a source image that can be searched for the earthquake location. We test the neural operators on both a synthetic dataset and a real dataset taken from the 2016 IRIS community wavefield experiment in Oklahoma \citep{sweet2018community}. We find that time reversal modeling with the trained neural operator enables us to locate earthquake locations from correlation-based source images with high resolution and efficiency. We use the 2D acoustic wave equations to illustrate our method but it is straightforward to extend the method to elastic and 3D cases. 

\section{Method}
We first briefly review time-reversal imaging of seismic sources. Then, we introduce the basics of neural operators and propose to achieve time reversal modeling by training a U-shaped neural operator for wave propagation with various source time functions. Our approach to synthesizing a training dataset for wavefield backpropagation is also detailed.

\subsection{Time reversal imaging}
The principle of time reversal states that time reversed wavefields will refocus to original source locations during backpropagation regardless of the complexity of the propagation medium since the acoustic wave equation in a nondissipative and heterogeneous medium is invariant for time reversal \citep{fink2006time}. Time reversal imaging with seismic data $d(\textbf{x}_s,\textbf{x}_r,t)$ emitted from passive sources $\textbf{x}_s$ and observed at stations $\textbf{x}_r$ consists of the following steps: 
\begin{itemize}
    \item Reversing seismograms in time with $d(\textbf{x}_s,\textbf{x}_r,T-t)=d(\textbf{x}_s,\textbf{x}_r,t)$ where $T$ denotes the time window; 
    \item With a predetermined velocity model, solving the wave equation with the reversed records $d(\textbf{x}_r,T-t)$ taken as new ``source time functions" excited at the receiver locations;
    \item Applying imaging conditions to the backpropagated wavefields $R_i, (i=1,...,N)$ where $N$ is the number of receivers;
    \item Searching for the spatial and temporal locations of sources $\textbf{x}_s$ in the imaging result $I(\textbf{x},t)$.
\end{itemize}

Conventional time reversal imaging employs the summation operator as an imaging condition:
\begin{equation}
I(\textbf{x},t) = \sum_{n=1}^{N} R_i (\textbf{x},t).  
\end{equation}
By searching for maximum intensities or best focusing of sources, we can obtain source images $I(\textbf{x},t_0)$ where $t_0$ denotes the origin time of the seismic event. The summation is in practice implicit since it is equivalent to backpropagating time-reversed records $d(\textbf{x}_r,T-t)$ at all receivers at once. Thus, only one simulation is required with the entire dataset. However, it surfers from low imaging resolution. Physical artifacts due to incoming wavefields from other sources and outgoing wavefields from focused sources make it challenging to search for source locations from low SNR dataset \citep{nakata2016reverse}. 
By contrast, correlation-based time reversal imaging considers one record at each receiver as an independent source for backpropagation, and replaces the summation operator by multiplication:   
\begin{equation}
I(\textbf{x},t) = \prod_{n=1}^{N} R_i (\textbf{x},t).
\end{equation}
The correlation imaging condition greatly improves the imaging quality by suppressing artifacts in conventional time reversal imaging, as they are supposed to be zero-valued in the resulting image. Ideally, hypocenters would only appear in $I(\textbf{x},t)$  when all the backpropagated wavefields coincide in both space and time. It turns out that this imaging condition is less sensitive to the frequency components of seismic data. However, treating each station individually is computationally demanding and thus challenging for real-time monitoring of seismicity. 

\subsection{Neural operators and U-shaped architecture}

Operator learning is an emerging machine learning paradigm that aims to learn mappings between function spaces. Recently, \cite{li2020neural,kovachki2021neural} proposed neural operators as data-driven grid-independent PDE solvers. The structure of a one-layer neural operator shares the formula of the general solution for linear PDEs parameterized by $\textbf{m}$:
\begin{equation}
u(\textbf{x}) = \int G_\textbf{m}(\textbf{x},\textbf{y})f(\textbf{y})d\textbf{y}.
\end{equation}
In seismology, $G$ denotes the Green's function of the wave equation, depending on $\textbf{m}$ (i.e. velocities discretized at $\textbf{x}$), and $f$ is the source time function. Likewise, neural operators use the following integral kernel operator as a basic form:
\begin{equation}
u(\textbf{x}) = (KV)(\textbf{x}) = \int K(\textbf{x},\textbf{y})V(\textbf{y})d\textbf{y} , 
\end{equation}
where $K$ denotes kernel convolution. By sequentially stacking $L$ linear kernel operators and introducing $L$ non-linear activation functions $\sigma$, we can build a neural operator $N_{\theta}$ parameterized by $\theta$ to estimate the solution of PDEs: 
\begin{equation}
N_{\theta} := Q \circ \sigma_{L}(W_{L}+K_L) \circ ... \circ \sigma_1(W_1+K_1) \circ P,
\end{equation}
where $W$ denotes a linear transform. $P$ is an encoder lifting the input to a high dimensional channel space. $Q$ is a decoder projecting the representation back to the original space. 

\begin{figure}
\centering
\includegraphics[width=\textwidth]{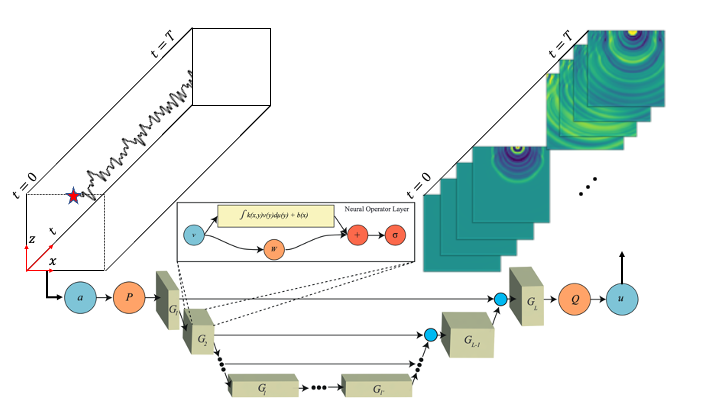}
\caption{U-shaped neural operator for time reversal modeling. The architecture contains $L$ Fourier neural layers: $G_i, i=1,...,L$. $P$ and $Q$ are up- and down-projections parameterized by neural networks, respectively. The input is a source time function $f(\textbf{x},t)$ sampled from a Gaussian random field. The output is the wavefield simulated with its input as source time function on a predetermined velocity model. The velocity model is not an implicit input but required for simulating a training dataset with a conventional PDE solver. For the 2D wave equation, both input and output are 3D volumes with equivalent number of grid points (2D space and 1D time).}
\label{fig:uno}
\end{figure}

Figure~\ref{fig:uno} shows the architecture of a U-shaped neural operator \citep[UNO;][]{rahman2022u} that we use in this study. By replacing the convolutional layers in the U-net \citep{ronneberger2015u} with Fourier neural operator layers \citep{li2020fourier}, UNO improves the memory performance of neural operators and allows for deep architecture. Here, each block $G_L$ represents a Fourier layer which uses Fourier transform $\mathcal{F}$ to compute the kernel convolution in the frequency domain:
\begin{equation}
(KV)(x) = \mathcal{F}^{-1}(R \cdot (\mathcal{F}(V)))(x), 
\end{equation}
where $R$ is the Fourier transform of a (learned) kernel function. These kernels can be efficiently computed by the Fast Fourier Transform, leading to an efficient solution of PDEs. 

To solve the efficiency problem of correlation-based time reversal imaging and make it suitable for real-time earthquake location, we employ neural operators as a fast and grid-independent solver for wave equations. As a functional space architecture, the method is independent of the discretization of velocity models. In other words, we can train a neural operator on low-resolution discretization and predict wavefields on fine grids, resulting in efficient high-resolution earthquake location.

\subsection{Training neural operators for time reversal modeling}

A neural operator has to be well trained to predict the solution of wave equations in an accurate manner. For time reversal modeling, we aim to train a neural operator to predict wavefields given arbitrary source time functions. If it generalizes well, the neural operator should be able to generate backpropagated wavefields for each station given any time-reversed seismogram as the input of the neural operator. 

\begin{figure}
\centering
\includegraphics[width=\textwidth]{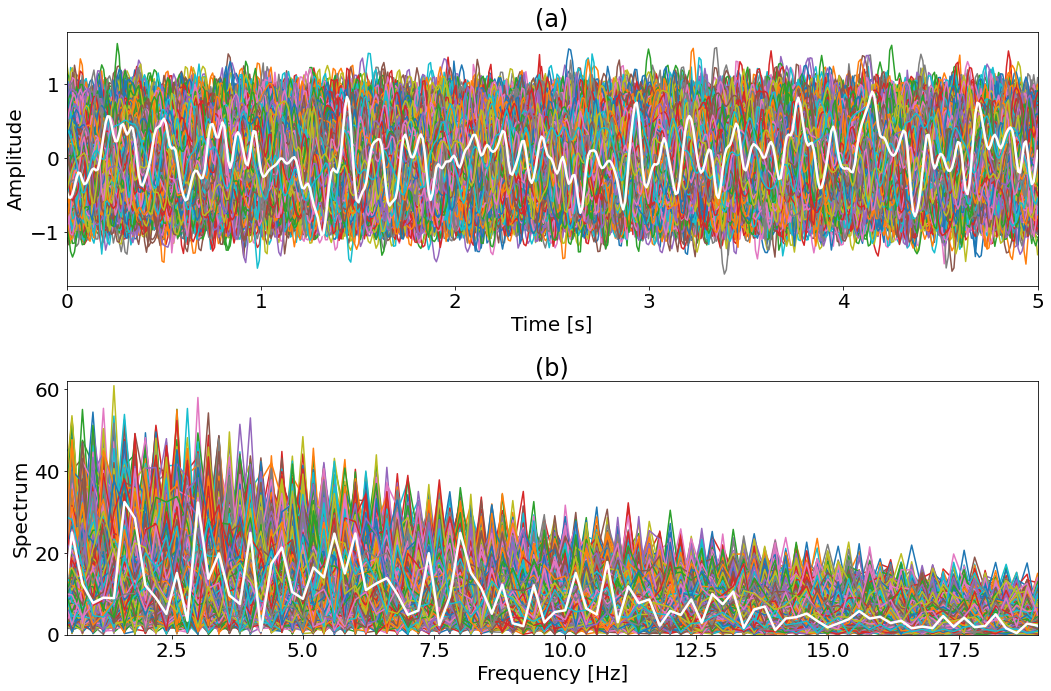}
\caption{Source time functions generated from a Gaussian random field for simulating the synthetic training and validation datasets. (a) Time series. (b) Frequency spectra. In total, we generate 2000 samples and use 1800 samples for training and 200 samples for validation. The white lines show one source time function and its frequency spectrum.}
\label{fig:randomfunction}
\end{figure}

To prepare such a training dataset, we sample random functions $f(t)$ from a Gaussian process $f \sim \mathcal{N}(0,k(t,t'))$ and use the random functions as source time functions for wave propagation. This is based mathematically on the fact that Gaussian processes are a dense subset of $L^2$ and can thus approximate any "physical" function to $\epsilon$ accuracy \citep{williams2006gaussian}. We choose a squared exponential covariance function as the kernel $k(t,t')$ of the Gaussian process:
\begin{equation}
k(t,t') = \sigma^2 exp(-\frac{1}{2} \frac{\parallel t-t' {\parallel}^2}{l^2}),
\end{equation}
where $\sigma^2$ and $l$ are hyperparameters denoting respectively a prior variance and correlation lengthscale. Considering each random function as a time series, $\sigma^2$ determines the amplitude, and $l$ controls its frequency spectrum. A small lengthscale generates high-frequency time series whereas low frequencies can be generated with a relatively large correlation lengthscale. With a prior variance of 0.1, we generate 2000 random functions with their correlation lengthscale randomly sampled from a uniform distribution: $l^2 \sim U[2,10]$ so that the dataset contains various frequency components (Figure~\ref{fig:randomfunction}). Each random function represents a time series with a recording time of 5 s and a sampling rate of 0.01 s.

\begin{figure}
\centering
\includegraphics[width=0.7\textwidth]{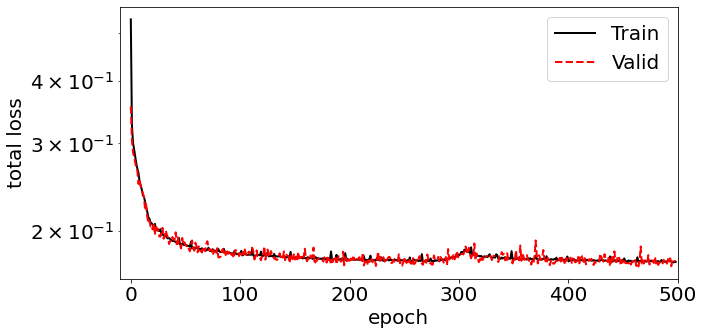}
\caption{Learning curves}
\label{fig:loss}
\end{figure}

\begin{figure}
\centering
\includegraphics[width=0.45\textwidth]{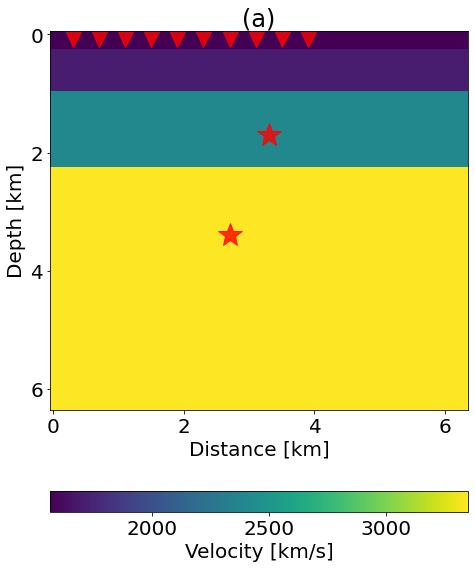}
\includegraphics[width=0.47\textwidth]{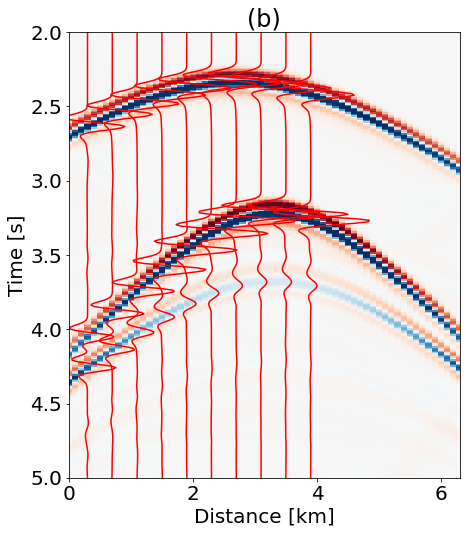}
\caption{(a) Acoustic velocity model (S-wave velocity model modified by \cite{fan2018investigating} for earthquake location at Oklahoma). Red stars indicate locations of seismic events. The source time functions are Ricker wavelet with different dominant frequencies (Tabel~\ref{tab:synthetic_sources}). These events are recorded by 64 stations evenly placed on the top of the model. Inverse triangles denote selected stations for studying the quality of time reversal imaging with a subset of receivers. (b) Seismograms. Red lines denotes ten traces recorded by the selected stations.} 
\label{fig:velocity}
\end{figure}

To prepare the output, we solve the 2D acoustic wave equation using spectral element modeling (SEM) on a predetermined acoustic wave velocity model (Figure~\ref{fig:velocity}). Although the velocity model is not an explicit input to the model, we need this prior knowledge to compute the wavefield $u(\textbf{x},t)$ with a conventional PDE solver when building a training dataset for a specific area. If trained well, the neural operator can be used to monitor and locate future earthquakes at a similar area. Here, the model is 6300 m in the vertical direction (depth) and $6300$ m in the horizontal direction (distance). With a grid spacing of 100 m, the computational domain has been discretized to $64\times64$. Each grid point on the surface represents a receiver location, so we have in total 64 stations in the acquisition geometry. With one random function as a source time function, each SEM simulation has a single source located at the surface of the velocity model. The horizontal location of the source in each simulation is random and sampled from a uniform distribution. In the end, the training dataset contains $M$ pairs of functions $\{f_i(\textbf{x},t),u_i(\textbf{x},t)\}_{i=1}^{M}$ with a 2D space discretization of $64\times64$ and 1D time discretization of $1\times500$. 

We train the neural operator from scratch using the Adam optimizer with a learning rate of $10^{-3}$ and a weight decay of $10^{-5}$. We separate 200 samples from the total 2000 simulations as a validation dataset. We observe that the neural operator did not benefit much from adding more training samples simulated in the same way, so we end up using 1800 samples for training. Both input and output are normalized before inputting into the neural operator by subtracting their means and dividing by their standard deviation. We use a hybrid loss function with both $\mathcal{L}_1$ and $\mathcal{L}_2$ losses:
\begin{equation}
\mathcal{L}_{hybrid} = \alpha \mathcal{L}_1 + \beta \mathcal{L}_2,  
\end{equation}
where $\alpha$ and $\beta$ are weights of the loss. Here we choose $\alpha=0.9$ and $\beta=0.1$ to train the neural operator. Figure~\ref{fig:loss} shows the training process with a batch size of one. The training takes approximately 135 hours for 500 epochs on a single NVIDIA Tesla V100 GPU. We stop training after 500 epochs and use the trained neural operator for time reversal modeling. The evaluation takes about 25 s for 64 individually backpropagated wavefields with a GPU memory usage of 8859 MB.  

\section{Results}
After training, the neural operator can be used for time reversal modeling and earthquake location. In this section, we evaluate the performance of the trained neural operator and report the results on one synthetic test dataset and one field dataset.  

\subsection{Synthetic data example}

\begin{figure}
\centering
\includegraphics[width=\textwidth]{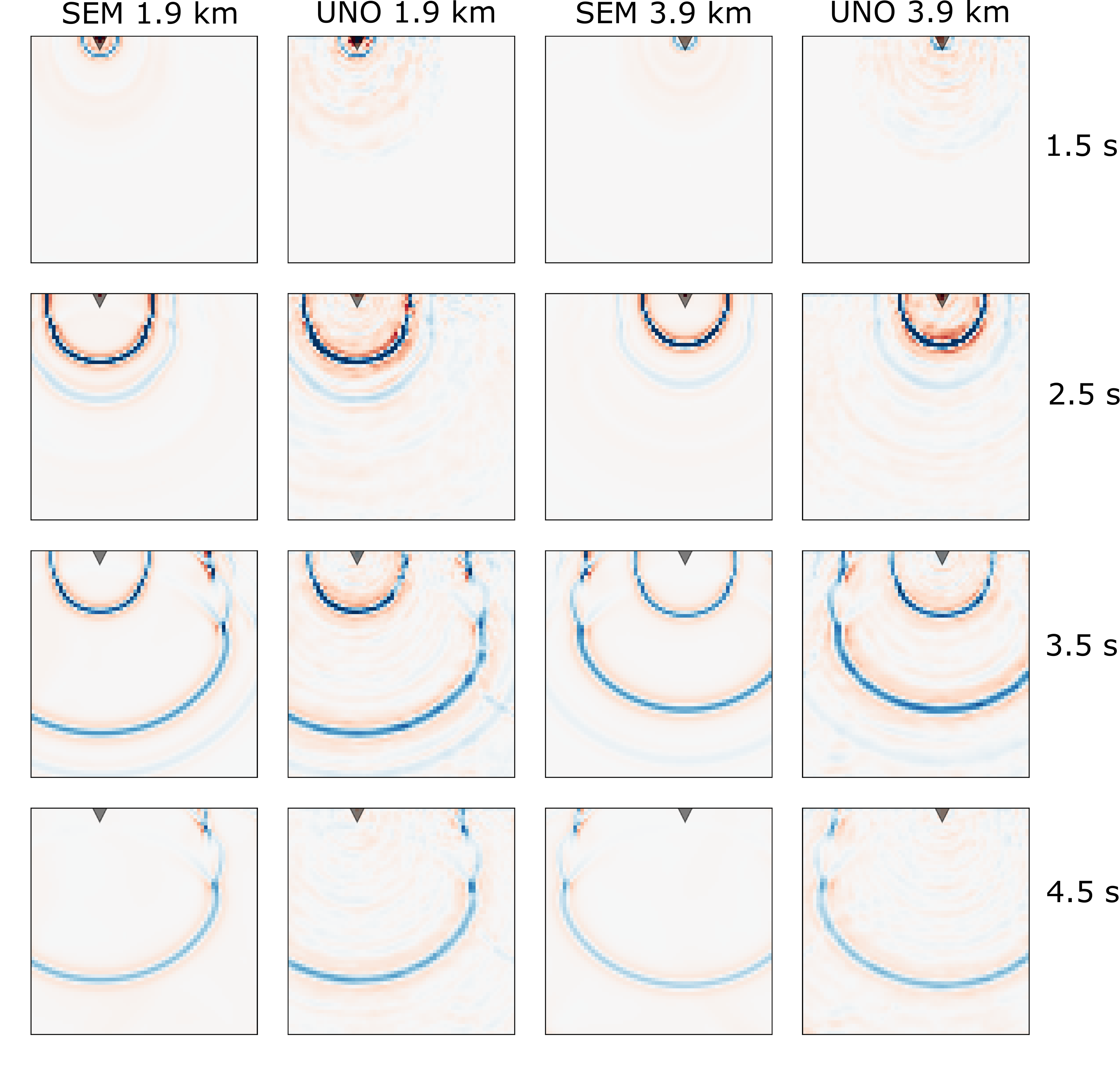}
\caption{Comparison of SEM and UNO wavefields for the synthetic test dataset at 1.5 s, 2.5 s, 3.5 s, and 4.5 s, respectively. Black triangles show locations of stations at 1.9 km and 3.9 km on the velocity model.}
\label{fig:synthetic_timereversal_wavefield}
\end{figure}

\begin{figure}
\centering
\includegraphics[width=\textwidth]{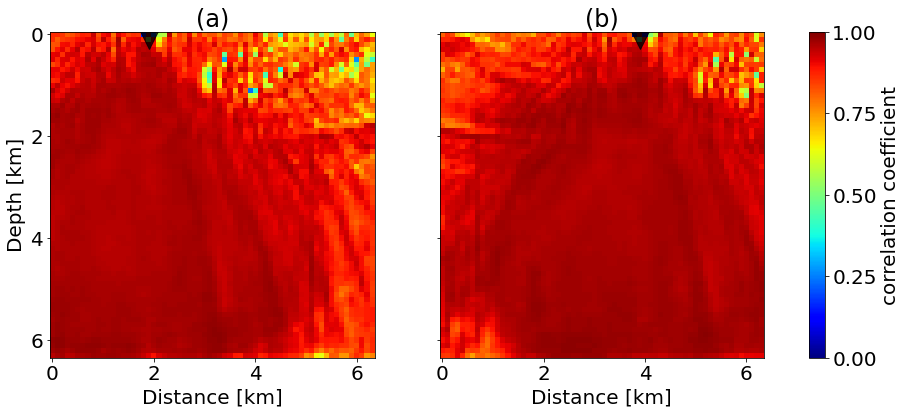}
\caption{Comparison of correlation coefficient between SEM and UNO wavefields at each grid point for the stations located at (a) 1.9 km and (b) 3.9 km, respectively. Black triangles show locations of stations.}
\label{fig:synthetic_timereversal_wavefield_cc}
\end{figure}

We first demonstrate the performance on a synthetic test dataset simulated on the predetermined velocity model (Figure~\ref{fig:velocity}a). This test includes two closely occurring events in the subsurface whose source time functions are Ricker wavelets with different dominant frequencies (Tabel~\ref{tab:synthetic_sources}). The time difference between two events is only 1.45 s, which is challenging for earthquake location due to the interference between them. Figure~\ref{fig:velocity}b shows the seismograms recorded by 64 stations on the surface. With their time-reversed seismograms as source time functions for time reversal modeling at corresponding receiver locations, Figure~\ref{fig:synthetic_timereversal_wavefield} compares the backpropagated wavefields computed using SEM and UNO at $T = 1.5$ s, $T=2.5$ s, $T=3.5$ s, and $T=4.5$ s for the stations located at 1.9 km and 3.9 km. Despite some amplitude discrepancies, the resulting UNO wavefields are very comparable with the reference SEM wavefields at different time steps, suggesting that the neural operator trained with random functions can generalize well and predict correct backpropagated wavefields for the synthetic events. Furthermore, Figure~\ref{fig:synthetic_timereversal_wavefield_cc} quantitatively evaluates the prediction by comparing the correlation coefficient of wavefields between SEM and UNO at each grid point. Although some mismatches exist at large offsets, the correlation coefficient is close to one at most of the nodes in the computational domain, implying the consistency between SEM and UNO for wave propagation. In particular, the trained UNO can generate a backpropagated wavefield for each receiver within several seconds, whereas SEM takes around 10 minutes for each simulation. 

\begin{table}
\caption{\label{tab:synthetic_sources}Source properties of the events in the synthetic example.}
\centering
\begin{tabular}{c c c c c}
\hline
 Event & Origin time [s] & Distance [km] & Depth [km] & Dominant frequency [Hz]  \\
\hline
  A & 2.32 & 3.30 & 1.70 & 7.00  \\
  B & 0.87 & 2.70 & 3.40 & 6.90  \\
\hline
\end{tabular}
\end{table}

\begin{figure}
\centering
\includegraphics[width=\textwidth]{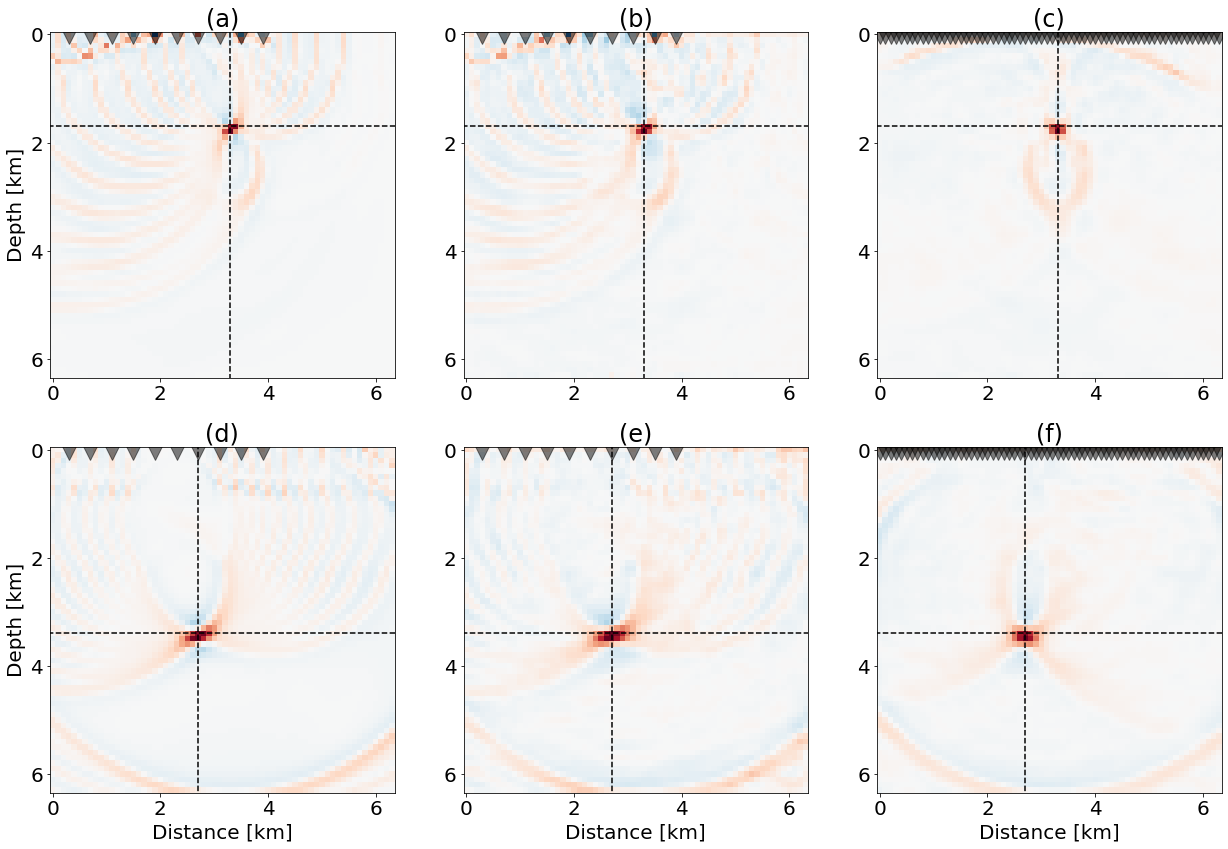}
\caption{Time reversal images produced by stacking (a and d) 10 SEM wavefields, (b and e) 10 UNO wavefields, and (c and f) 64 UNO wavefields. Comparison of the source images between (a-c) the shallower event and (d-f) the deeper event. Black triangles and crossings of dash lines show the locations of receivers and sources, respectively.}
\label{fig:time_reversal_image_stacking}
\end{figure}

We then locate the events with time reversal imaging using the resulting backpropagated wavefields. Figure~\ref{fig:time_reversal_image_stacking} compares time reversal images by stacking (Equation 1) 10 SEM wavefields and 10 UNO wavefields from the selected stations. Due to the accuracy of UNO for wave propagation, the difference of source images between SEM and UNO wavefields is minimal, but wave propagation with UNO is much faster than the conventional SEM solver. All images stacked with 10 wavefields show aperture artifacts (patterns of intersecting rings) due to limited aperture acquisition. These aperture artifacts can be effectively suppressed by stacking all receiver wavefields (Figure~\ref{fig:time_reversal_image_stacking} c and f). However, the interference from one event does not diminish when locating the other event with the stacking imaging condition.  

\begin{figure}
\centering
\includegraphics[width=\textwidth]{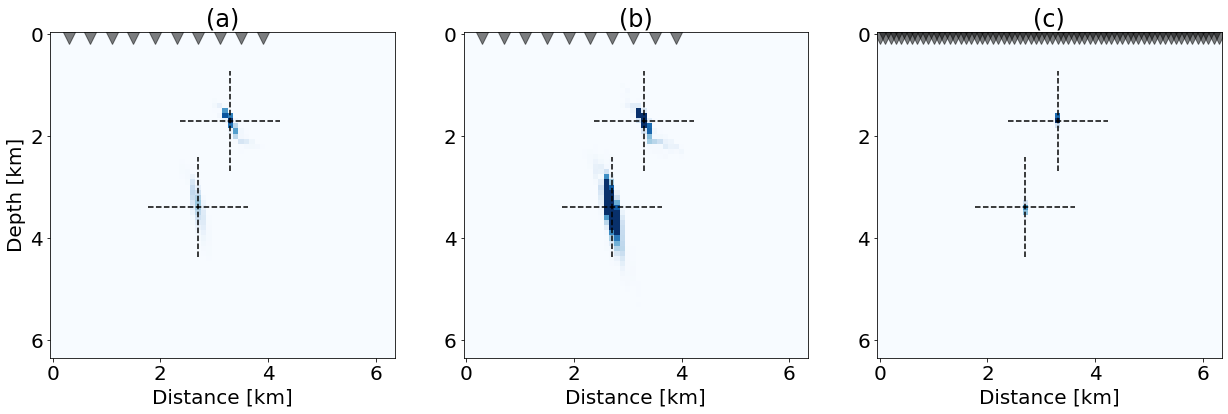}
\caption{Time reversal images produced by correlating (a) 10 SEM wavefields, (b) 10 UNO wavefields, and (c) 64 UNO wavefields. Black triangles and crossings of dash lines show the locations of receivers and sources, respectively.}
\label{fig:time_reversal_image_correlating}
\end{figure}

Figure~\ref{fig:time_reversal_image_correlating} shows time reversal images by correlating backpropagated wavefields (Equation 2). Artifacts exist with only 10 receiver wavefields. But correlation with 64 wavefields results in a clean source image with much higher resolution compared with the results by stacking. The artifacts due to a sparse receiver array are also reduced when correlating more wavefields, implying the importance of dense seismic network for correlation-based time reversal imaging. With neural operators, we can efficiently calculate backpropagated wavefields for each receiver in a seismic network and achieve real-time earthquake location with high resolution.

\subsection{Field data example}

\begin{figure}
\centering
\includegraphics[width=\textwidth]{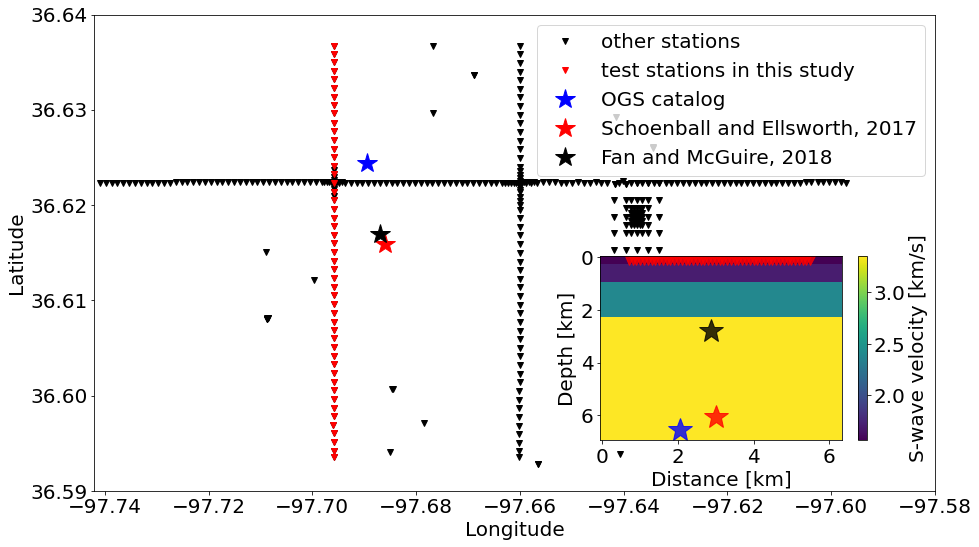}
\caption{2016 IRIS community wavefield experiment in Oklahoma. Blue, red, and black stars indicate the hypocenter of the 07/11/2016 M2 earthquake reported by Oklahoma Geological Survey (OGS), \cite{schoenball2017waveform}, and \cite{fan2018investigating}, respectively. Red triangles show the line of stations used in this study for locating the earthquake with time reversal imaging. The lower right insert shows the S-wave velocity model, the reported locations of the earthquake, and the relative positions of the selected stations on the velocity model.}
\label{fig:field_geometry}
\end{figure}

We test the trained neural operator on the field data collected during 2016 IRIS community wavefield experiment in Oklahoma \citep{sweet2018community}. Since the training dataset is simulated on the local S-wave velocity model, we can directly locate earthquakes for the field data with the trained neural operator. The seismic network mainly contains $\sim$ 250 node stations deployed in three lines: one east-west line and two north-south lines (Figure~\ref{fig:field_geometry}). One M2 earthquake occurred on July 11, 2016 during the nodal deployment. We use SH waves recorded by the eastern channel of the western north-south line to locate the earthquake. By assuming an acoustic medium, we backpropagate SH waves on the S-wave velocity model and solve 2D acoustic wave equation for wave propagation. 

\begin{figure}
\centering
\includegraphics[width=\textwidth]{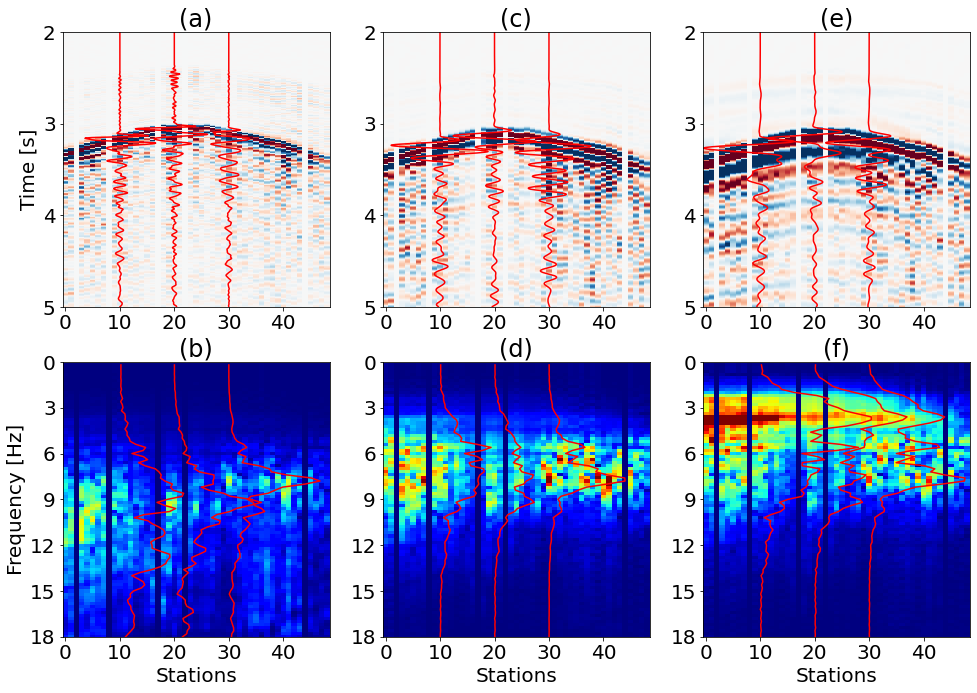}
\caption{The E-component of seismograms recorded by the selected stations on the eastern north-south line. The station interval is around 100 m. (a) Raw waveforms without preprocessing and (b) their frequency spectra. (c) Waveforms filtered by a lowpass filter with a corner frequency of 6 Hz and (d) their frequency spectra. (e) Waveforms filtered by a lowpass filter with a corner frequency of 3 Hz and (f) their frequency spectra. Red lines show waveforms and spectra of the stations located at 1.7 km, 2.7 km, and 3.7 km on the velocity model (Figure~\ref{fig:field_geometry}).}
\label{fig:field_records}
\end{figure}

Figure~\ref{fig:field_geometry} compares the location of the M2 earthquake determined with various methods. As reported by Oklahoma Geological Survey, the depth of the M2 earthquake is 6.562 km. \cite{schoenball2017waveform} relocate earthquakes at Oklahoma and Southern Kansas using a pick-based earthquake location workflow and determine that the depth is about 6.051 km. \cite{fan2018investigating} locate the earthquake by fitting the S–P times at each station of the three lines with 3D grid search and report the earthquake location at latitude $36.617^{\circ}$, longitude $97.687^{\circ}$, and depth 2.8 km. We find that our location result using the single line is very close to that determined by \cite{fan2018investigating}. We therefore choose to use the result by \cite{fan2018investigating} as the reference earthquake location. However, it should be noted that our result will not be exactly at the reference location due to the 2D assumption in this study.

We select 49 stations on the line with an approximately uniform interval of 100 m. Figure~\ref{fig:field_records} shows 5-s long waveforms and their low-frequency components lowpass filtered with corner frequencies of 6 Hz and 3 Hz, respectively. Five of the stations are dead, so we use the left 44 stations for time reversal imaging. With different frequency components for backpropagation, we could compare the influence of bandwidth on wave propagation and time reversal imaging with neural operators.

\begin{figure}
\centering
\includegraphics[width=\textwidth]{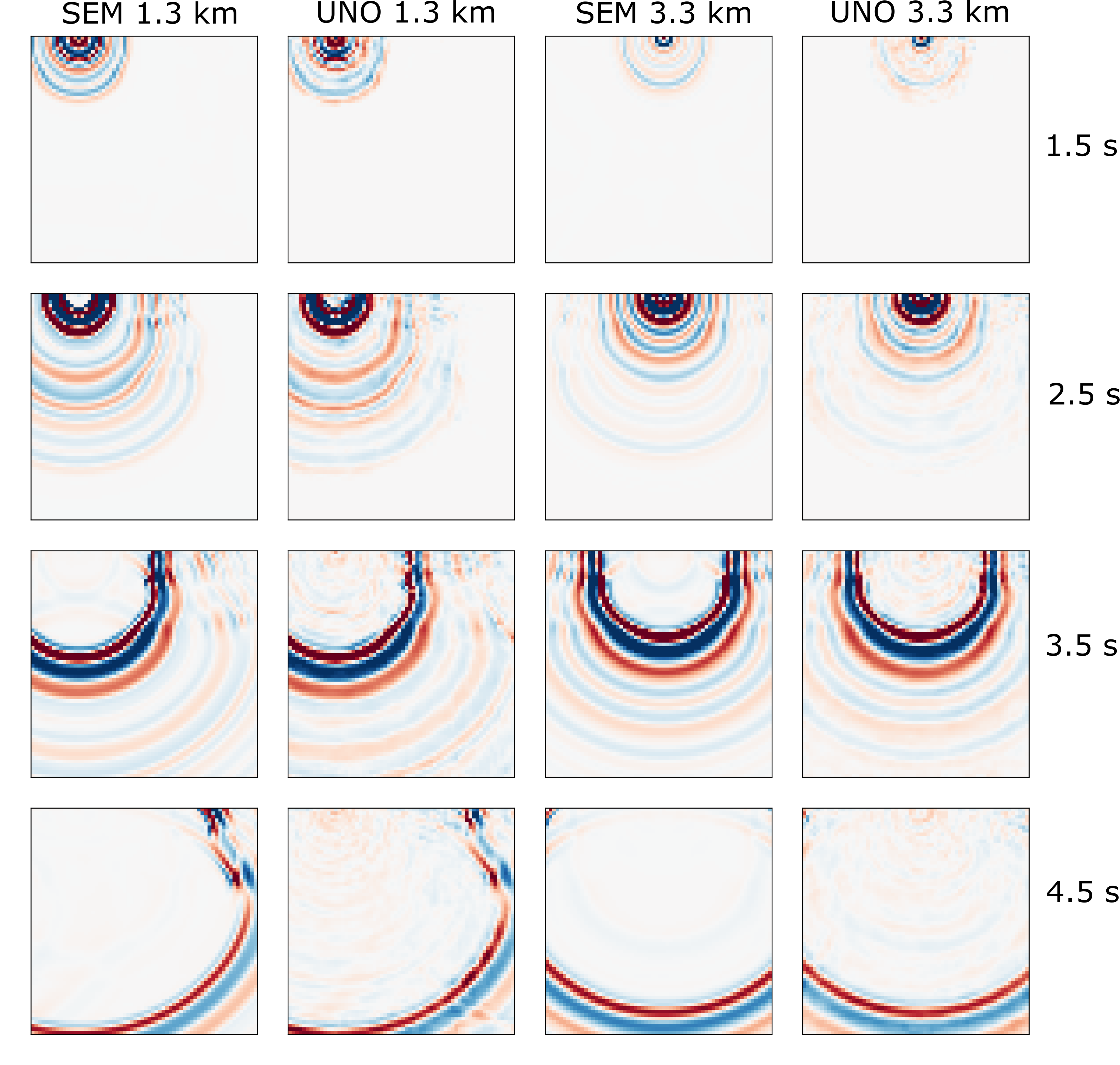}
\caption{Backpropagated wavefields for the field data calculated by SEM and UNO at 1.5 s, 2.5 s, 3.5 s, and 4.5 s, respectively. The stations are located at 1.3 km and 3.3 km, respectively. The inputs of SEM and UNO are seismograms processed by a low-pass filter with a corner frequency of 3 Hz.}
\label{fig:ok_wavefield_3hz}
\end{figure}

\begin{figure}
\centering
\includegraphics[width=\textwidth]{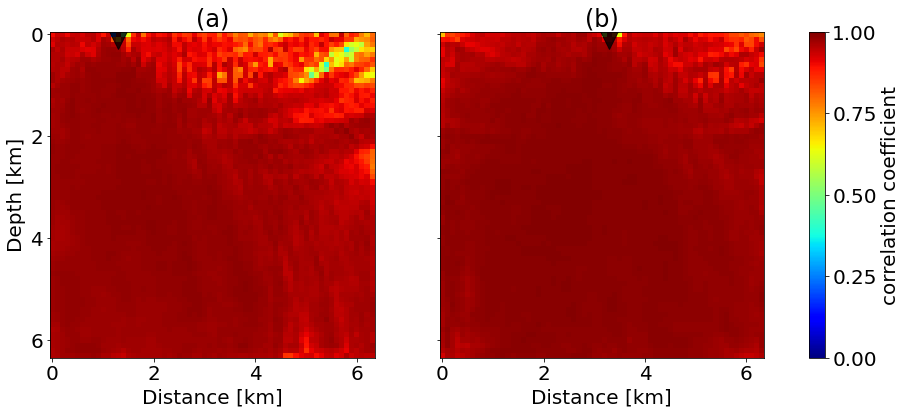}
\caption{Correlation coefficient between SEM and UNO wavefields for the stations located at (a) 1.3 km and (b) 3.3 km, respectively. Black triangles show locations of stations. The inputs of SEM and UNO are seismograms processed by a low-pass filter with a corner frequency of 3 Hz. }
\label{fig:event_wavefield_3hz_cc}
\end{figure}

\begin{figure}
\centering
\includegraphics[width=\textwidth]{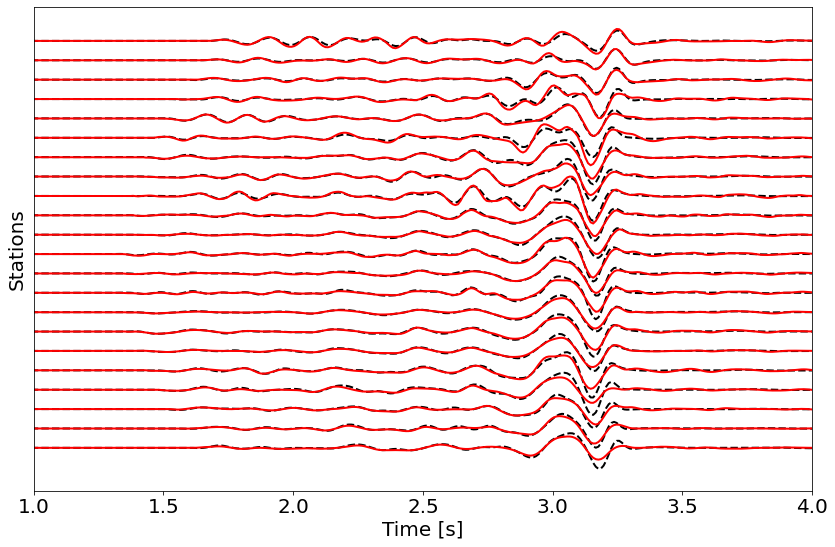}
\caption{Backpropagated waveforms recorded at the reference earthquake location during time reversal modeling of the field data. Black dash line and red solid line denote the results of SEM and UNO, respectively. In total, 22 out of the 45 stations are compared.}
\label{fig:ok_wavefield_3hz_trace}
\end{figure}

\begin{figure}
\centering
\includegraphics[width=\textwidth]{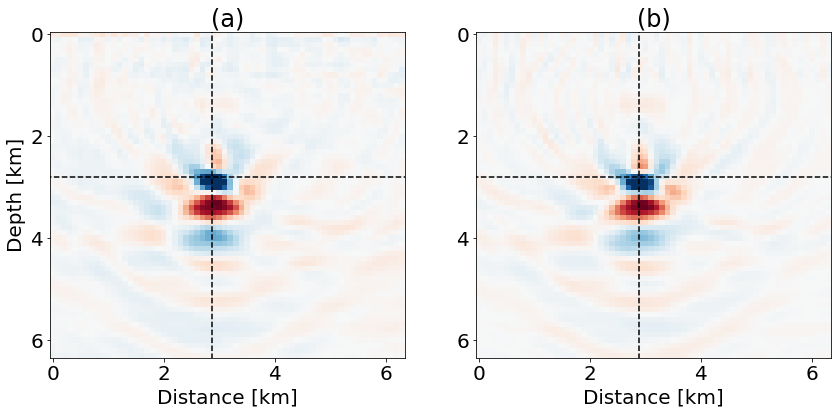}
\includegraphics[width=\textwidth]{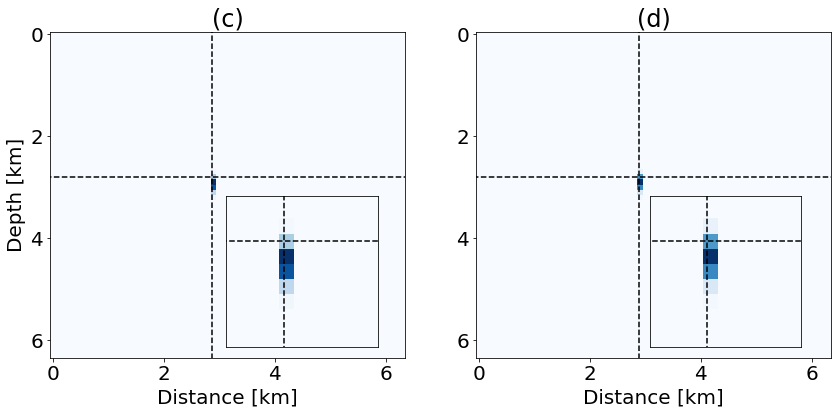}
\caption{Comparison of time reversal images produced by (a) stacking UNO wavefields, (b) stacking SEM wavefields, (c) correlating UNO wavefields, and (d) correlating SEM wavefields. The input seismograms for backpropagation have been processed by a low-pass filter with a corner frequency of 3 Hz.}
\label{fig:ok_wavefield_3hz_image}
\end{figure}

\begin{figure}
\centering
\includegraphics[width=\textwidth]{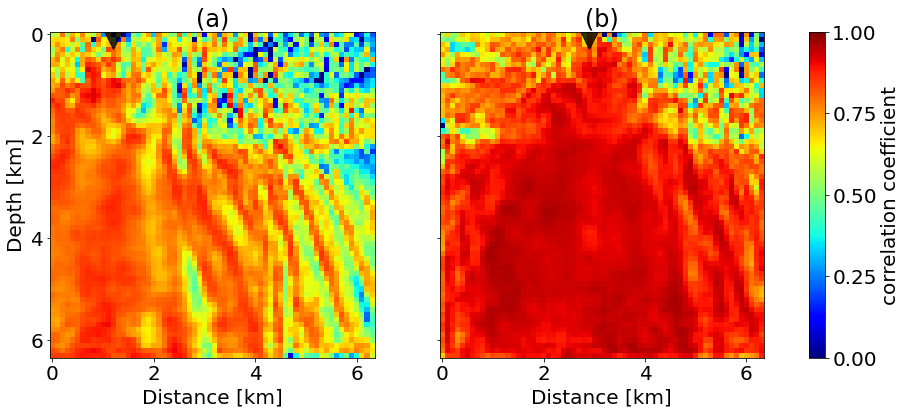}
\caption{Correlation coefficient between SEM and UNO wavefields for the stations located at (a) 1.3 km and (b) 3.3 km, respectively. Black triangles show locations of stations. The input seismograms of SEM and UNO have been processed by a low-pass filter with a corner frequency of 6 Hz.}
\label{fig:event_wavefield_6hz_cc}
\end{figure}

\begin{figure}
\centering
\includegraphics[width=\textwidth]{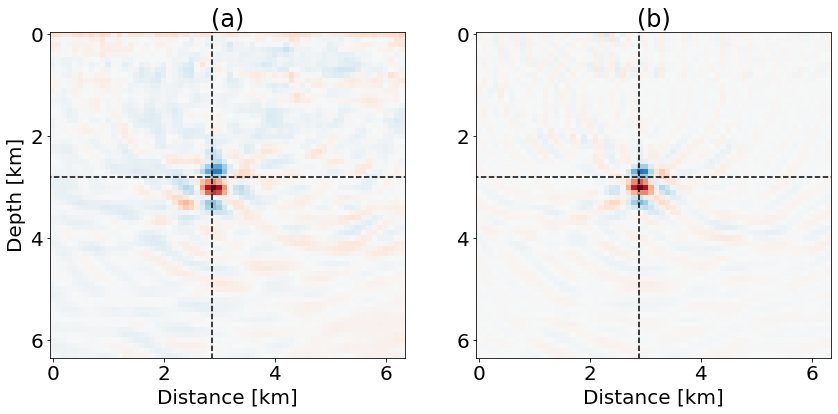}
\includegraphics[width=\textwidth]{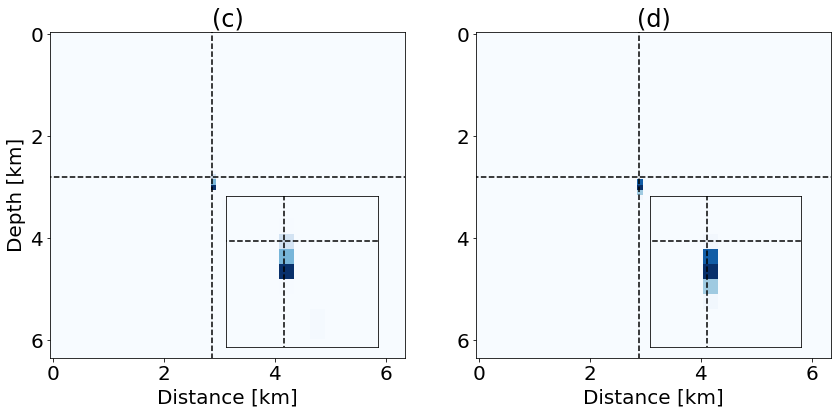}
\caption{Comparison of time reversal images produced by (a) stacking UNO wavefields, (b) stacking SEM wavefields, (c) correlating UNO wavefields, and (d) correlating SEM wavefields. The input seismograms for backpropagation have been processed by a low-pass filter with a corner frequency of 6 Hz. }
\label{fig:ok_wavefield_6hz_image}
\end{figure}

\begin{figure}
\centering
\includegraphics[width=\textwidth]{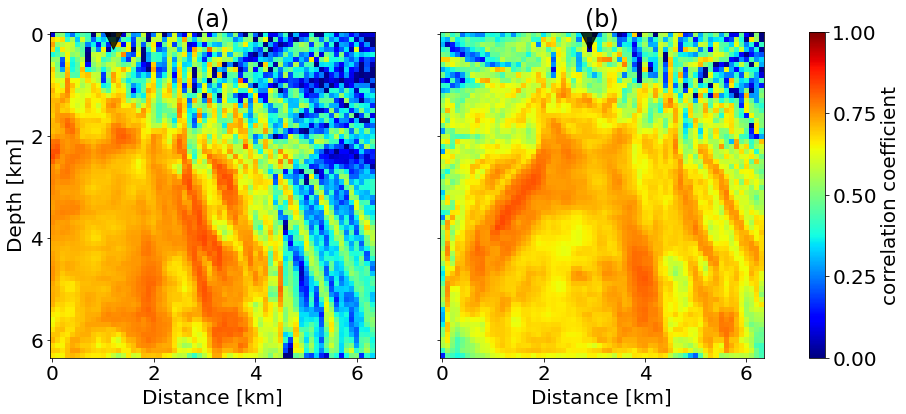}
\caption{Comparison of time reversal images produced by (a) stacking UNO wavefields, (b) stacking SEM wavefields, (c) correlating UNO wavefields, and (d) correlating SEM wavefields. The inputs of SEM and UNO are raw seismic waveforms of the field data without any pre-processing.}
\label{fig:event_wavefield_raw_cc}
\end{figure}

\begin{figure}
\centering
\includegraphics[width=\textwidth]{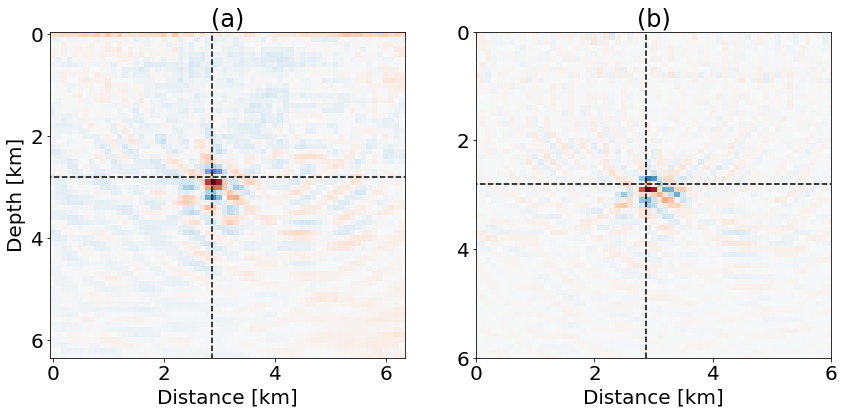}
\includegraphics[width=\textwidth]{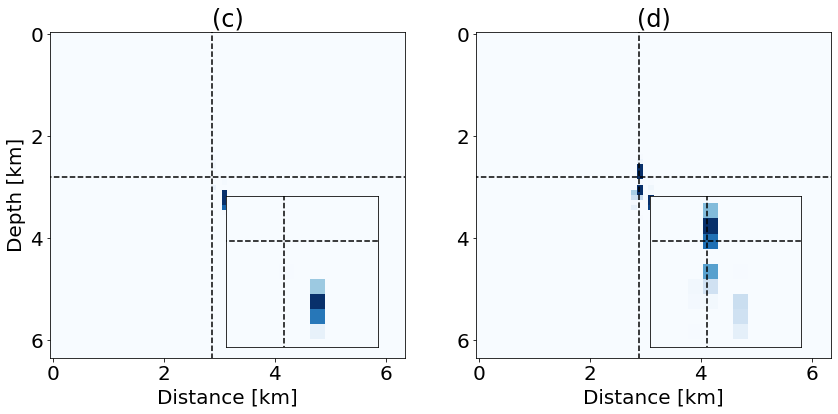}
\caption{Comparison of time reversal images produced by (a) stacking UNO wavefields, (b) stacking SEM wavefields, (c) correlating UNO wavefields, and (d) correlating SEM wavefields. The inputs of SEM and UNO are raw seismic waveforms of the field data without any pre-processing.}
\label{fig:ok_wavefield_raw_image}
\end{figure}

\begin{figure}
\centering
\includegraphics[width=\textwidth]{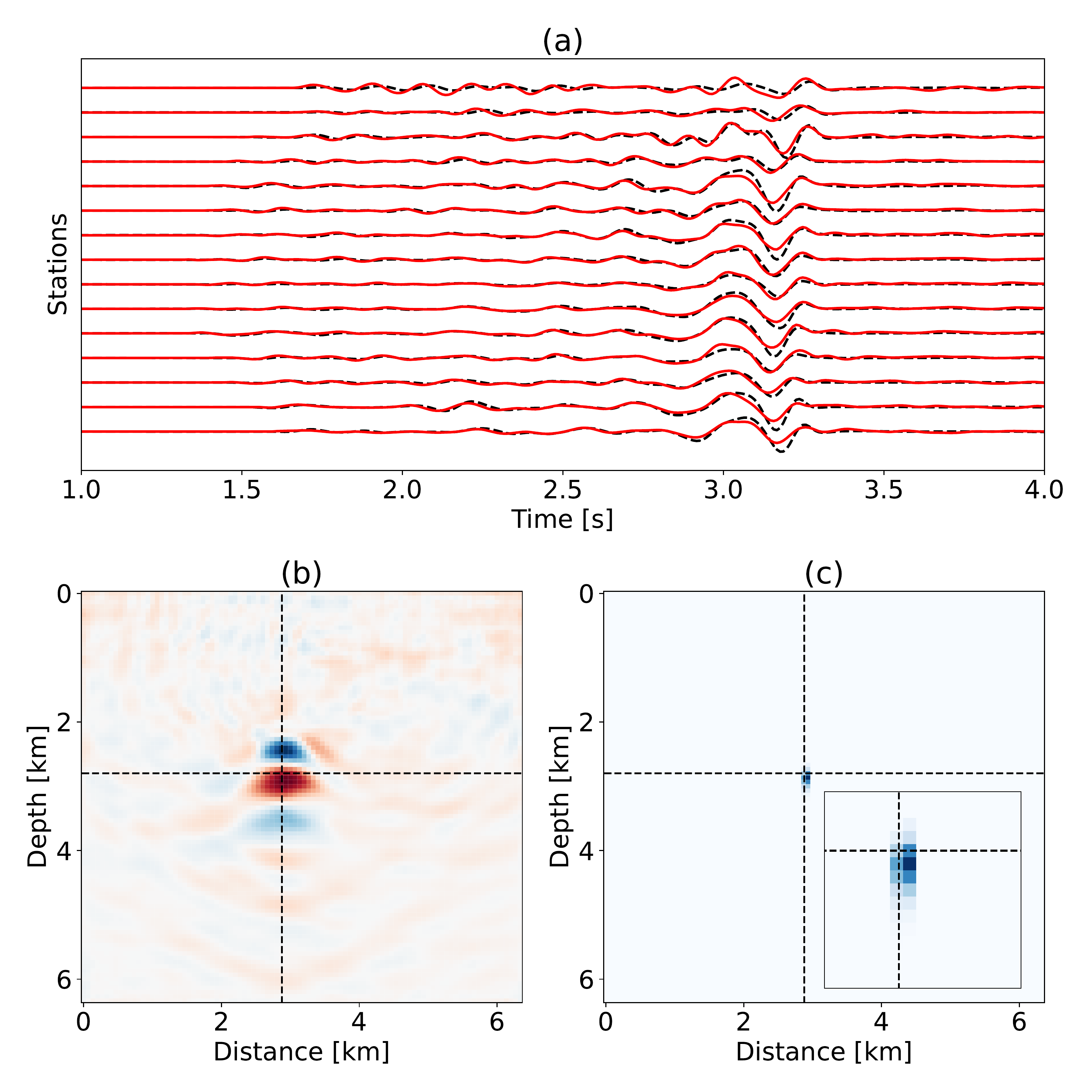}
\caption{Modeling and imaging results for spatial resolution $96\times96$ using the neural operator trained on $64\times64$. (a) Backpropagated waveforms recorded at the reference earthquake location for different stations. In total, 15 out of the 45 stations are compared. Black dash line and red solid line denote SEM and UNO, respectively. Time reversal images by (b) stacking and (c) correlating UNO wavefields. The inputs of SEM and UNO are the waveforms filtered by a lowpass filter with a corner frequency of 3 Hz.}
\label{fig:highres}
\end{figure}

Using each time-reversed trace in Figure~\ref{fig:field_records} as a source time function, we individually backpropagate records of the 44 stations with SEM and the trained neural operator. Figure~\ref{fig:ok_wavefield_3hz} compares the UNO and SEM wavefields simulated with the 3 Hz lowpass filtered seismograms at four timesteps: 1.5 s, 2.5 s, 3.5 s, and 4.5 s for the stations located at 1.3 km and 3.3 km. The UNO trained with a synthetic dataset successfully generalizes to the field data and correctly predicts the backpropagation process, except for some errors near the edges of the computational domain. The correlation coefficient confirms the high accuracy of UNO for wave propagation compared with SEM (Figure~\ref{fig:event_wavefield_3hz_cc}). Figure~\ref{fig:ok_wavefield_3hz_trace} compares the backpropagated waveforms recorded at the reference earthquake location during time reversal modeling for each station. We observe that the waveform difference between UNO and SEM is very small. At the reference earthquake location, waveforms from all stations show a maximum absolute amplitude at around 3.2 s. By either stacking or correlating these wavefields, the maximum intensity is enhanced and thus allows us to determine the occurring time and location of the earthquake. 

Figure~\ref{fig:ok_wavefield_3hz_image} compares time reversal images using 3 Hz lowpass filtered data from all 44 stations. With the high accuracy of the UNO for wave propagation, we obtain very similar source images with either UNO or SEM wavefields. Time reversal imaging by stacking shows focused energy with sidelobes around the reference earthquake location and exhibit a large width of focusing due to the long wavelengths. The correlation image has high-resolution spike-like source images and allows us to locate the earthquake accurately. We find that the focusing coincides with the reference earthquake location with minor shifts, which confirms the effectiveness of our approach. 

We further study the performance of UNO when dealing with high-frequency seismograms. Figure~\ref{fig:event_wavefield_6hz_cc} compares the correlation coefficient between wavefields backpropagated by UNO and SEM using seismograms filtered by a 6 Hz lowpass filter. With higher frequencies, UNO is not as accurate but the resulting time reversal images are very comparable, in particular with a correlation-based imaging condition (Figure~\ref{fig:ok_wavefield_6hz_image}).

Figure~\ref{fig:event_wavefield_raw_cc} shows the correlation coefficient between the wavefields backpropagated by UNO and SEM without any preprocessing on the time-reversed seismograms. Compared with the results with lower frequencies as the input (Figures~\ref{fig:event_wavefield_3hz_cc} and \ref{fig:event_wavefield_6hz_cc}), UNO has larger errors at far offsets and near surfaces. Since the frequency components of the raw seismograms are much higher than those of the synthetic training dataset, it is challenging for UNO to predict the fullband wavefields correctly. However, we observe that the event waveforms are predicted at correct positions in the backpropagated wavefields at different time steps, which is the main information relied on to locate the earthquake. Stacking the UNO wavefields produces a time reversal image with lower frequency components compared with the result using SEM wavefields (Figure~\ref{fig:ok_wavefield_raw_image}). Nevertheless, high-frequency noise is not backpropagated in UNO wavefields, resulting in a cleaner correlation-based source image compared with the image with SEM wavefields.

The trained UNO can evaluate solutions of the wave equation at different discretizations of the velocity model because they are mesh-invariant. Consequently, we can obtain a high-resolution image of sources by backpropagating wavefields at a fine discretization using the UNO trained with a coarse gird. Figure~\ref{fig:highres} shows the modeling and imaging results of UNO evaluated at $96\times96$ but trained at $64\times64$ discretization. Although the evaluation resolution is 1.5 times higher than the training resolution, the UNO can successfully predict the wavefields which are comparable with SEM wavefields. The finer discretization enables us to determine source locations from correlation-based source images with less uncertainty and equivalent computational cost. On the contrary, computational cost of conventional PDE solvers increases dramatically with increasing number of grid points in the computational domain. Tabel~\ref{tab:compuationtime} compares the computational time of 45 simulations between UNO and SEM. SEM is very time-consuming and takes about five hours for all simulations. With UNO, we can predict the backpropagated wavefields for all stations within one minute, which enables us to locate earthquakes with high resolution and in a real-time fashion.

\begin{table}
\caption{\label{tab:compuationtime}Computational time of UNO and SEM for the field dataset (45 simulations).}
\centering
\begin{tabular}{c c c}
\hline
 Discretization & UNO & SEM  \\
\hline
  $64 \times 64$ & 25 s & 3 hours  \\
  $96 \times 96$ & 28 s & 9 hours  \\
\hline
\end{tabular}
\end{table}

\section{Discussion and limitations}

In order to accelerate time reversal modeling, we train neural operators for seismic wave propagation with various source time functions. The trained neural operator can predict time-reversed wavefields at all time-steps simultaneously for each station within negligible time, which greatly accelerates the expensive wave propagation process of time reversal imaging. Thanks to the efficiency of neural operators and high spatiotemporal resolution of the correlation imaging condition, we may achieve real-time and high-resolution earthquake location and monitoring directly from multi-station waveforms. 

The performance of UNO depends on the quality and properties of the training dataset, and hence, the frequency spectra of the source time functions in the training dataset should be comparable to the seismograms used for earthquake location. In our field-data example, UNO provides more accurate wavefields when dealing with the seismograms filtered by a lowpass filter with a corner frequency of 6 Hz and 3 Hz, respectively. Since our predefined training dataset is dominated by functions with relatively low frequencies, directly feeding raw seismograms to the trained neural operators predicts mainly low-frequency components. However, these low frequencies seem to be sufficient to locate the earthquakes with the time reversal approach. If it is desirable to expand this frequency range, then the SEM simulations should be performed with random source time functions having these properties. In addition, training datasets are generally simulated with conventional wave-equation solvers. Artifacts such as numerical dispersion and boundary reflections should be avoided; otherwise, neural operators may treat these artifacts as signals when learning wave propagation. 

Furthermore, our approach is designed to locate and monitor earthquakes in a specific area. Thus, prior knowledge of the velocity structure for that area is required for simulating a training dataset. Similar to other time reversal methods, the uncertainty of velocity models will affect the accuracy of earthquake location. Another limitation comes from the geometry of the seismic network. Using only surface acquisitions will limit spatial resolution in the vertical direction, which has been observed in the imaging results in both examples. 

We focus on the 2D acoustic wave equation in the present work but the extension of the work to more complex media, such as elasticity and attenuation is possible provided the neural operator is redesigned and trained with relevant wave equations. The benefits of neural operators in fast wave propagation should be more obvious for elastic wave equations. Moreover, extending the work to realistic earthquake location problems may require a memory-efficient neural operator for solving the 3D wave equation, which is left for future study. 

\section{Conclusion}
We present a deep-learning approach for real-time earthquake location with correlation-based time reversal imaging. The method leverages a U-shaped neural operator for seismic wave propagation, which enables us to calculate a back-propagated wavefield for each station within negligible time and satisfactory accuracy. Unlike pick-based earthquake location workflow, our method determines earthquake hypocenters directly with full waveforms recorded by multiple stations. We evaluate the method with both synthetic and field examples. The accuracy of the UNO trained with a Gaussian random field is very comparable with SEM when testing on field data, in particular for low-frequency waveforms. Although backpropagating raw seismograms degrades the performance of time reversal modeling, the correlation imaging condition seems to be robust with amplitude errors and results in a source image with correct earthquake locations. With a simple preprocessing step (lowpass filter), the method has potential for real-time earthquake location and monitoring. The extension of the method to elastic media and 3D models is straightforward, as long as we modify neural operators for wave propagation in these scenarios. 

\section{Acknowledgement}
H. Sun thanks Tong Bai for his helpful discussion. The field data are available at IRIS Data Management Center under network code YW and can be downloaded at \url{http://ds.iris.edu/mda/YW/?timewindow=2016-2016}.

\bibliographystyle{seg}
\bibliography{main}

\end{document}